\documentclass[conference]{IEEEtran}
\IEEEoverridecommandlockouts
\usepackage{cite}
\usepackage{amsmath,amssymb,amsfonts, amsthm}
\usepackage{algorithmic}
\usepackage{graphicx}
\usepackage{textcomp}
\usepackage{xcolor}
\usepackage{comment}
\def\BibTeX{{\rm B\kern-.05em{\sc i\kern-.025em b}\kern-.08em
    T\kern-.1667em\lower.7ex\hbox{E}\kern-.125emX}}
\begin{document}

\title{HashCore: Proof-of-Work Functions for General Purpose Processors
}

\author{\IEEEauthorblockN{Yanni Georghiades}
\IEEEauthorblockA{\textit{Department of ECE} \\
\textit{The University of Texas at Austin}\\
Austin, United States \\
yanni.georghiades@utexas.edu}
\and
\IEEEauthorblockN{Steven Flolid}
\IEEEauthorblockA{\textit{Department of ECE } \\
\textit{The University of Texas at Austin}\\
Austin, United States \\
stevenflolid@utexas.edu}
\and
\IEEEauthorblockN{Sriram Vishwanath}
\IEEEauthorblockA{\textit{Department of ECE} \\
\textit{The University of Texas at Austin}\\
Austin, United States \\
sriram@utexas.edu}
}

\maketitle

\begin{abstract}
Over the past five years, the rewards associated with mining Proof-of-Work blockchains have increased substantially. 
As a result, miners are heavily incentivized to design and utilize Application Specific Integrated Circuits (ASICs) that can compute hashes far more efficiently than existing general purpose hardware. 
Currently, it is difficult for most users to purchase and operate ASICs due to pricing and availability constraints, resulting in a relatively small number of miners with respect to total user base for most popular cryptocurrencies. 
In this work, we aim to invert the problem of ASIC development by constructing a Proof-of-Work function for which an existing general purpose processor (GPP, such as an x86 IC) is {\em already an optimized ASIC}. 
In doing so, we will ensure that any would-be miner either already owns an ASIC for the Proof-of-Work system they wish to participate in or can attain one at a competitive price with relative ease. 
In order to achieve this, we present {\bf HashCore}, a Proof-of-Work function composed of ``widgets" generated pseudo-randomly at runtime that each execute a sequence of general purpose processor instructions designed to stress the computational resources of such a GPP. 
The widgets will be modeled after workloads that GPPs have been optimized for, for example, the SPEC CPU 2017 benchmark suite for x86 ICs, in a technique we refer to as {\em inverted benchmarking}. 
We provide a proof that HashCore is collision-resistant regardless of how the widgets are implemented.
We observe that GPP designers/developers essentially  create an ASIC for benchmarks such as SPEC CPU 2017. 
By modeling HashCore after such benchmarks, we create a Proof-of-Work function that can be run most efficiently on a GPP, resulting in a more accessible, competitive, and balanced mining market.

\end{abstract}



\section{Introduction}
\label{sec:intro}

In recent years, Proof-of-Work protocols have grown in popularity in large part due to their application in cryptocurrencies such as Bitcoin or Ethereum. 
While decentralized cryptocurrencies exist in many varieties, one common feature is the use of a blockchain as a tamper-evident distributed ledger. 
In order to discourage malicious actors from attempting to overwrite the ledger, many cryptocurrencies employ a Proof-of-Work (PoW) protocol, defined first in \cite{pow:dwork} and formalized in \cite{pow}.

In the context of blockchains, PoW protocols simply require that the header for each block can be passed through a hash function such that the resulting hash meets some statistically unlikely structural requirement, such as some number of leading zeros in its binary representation. 
We call a hash meeting this requirement a ``proof of work'' because we know with high probability that sufficient computational ``work'' must have been performed to find this hash. 
This ensures that on average, the cost of overwriting blocks (``pages" in the ledger) will be comparable to the cost of initially writing those blocks.
As most PoW systems vary the difficulty of the PoW protocol with the total hashing power of the network, we have the important property that an increase in hashing power results in an increased cost of attacking the network. 
The process of searching for hashes is referred to as ``mining,'' and there is generally a significant reward associated with successfully mining a block. 

Given the competitive nature of mining, each miner is heavily incentivized to increase the efficiency with which they can compute hashes. 
A common strategy is to develop or purchase custom hardware in the form of an Application Specific Integrated Circuit (ASIC) that can efficiently execute the hash function used by the PoW protocol. 
ASICs are generally much more cost efficient than generalized hardware and are currently the best way to profitably mine most popular cryptocurrencies. 
This is acceptable for a PoW system so long as such ASICs are widely accessible to a large number of users for roughly the same price. 
Unfortunately, this is not the case, as ASIC development is a highly expensive and time-investive process. 
Thus, access to an ASIC is restricted to those that can afford them. 
For this reason, cryptocurrency mining has tended toward centralization in the recent past, with the total hashing power of the network being controlled by a few large mining operations rather than being diffused across the much larger user base as would be desirable in a truly decentralized system. 
Our proposed solution to this problem is to develop a new {\em PoW function}  for which the average user already owns an optimized-ASIC, thereby reducing the barrier to entering the mining market and increasing the availability of such optimized ASICs to any user.

In previous work, the most common approach for designing a PoW function that cannot be optimized through custom hardware has been targeting a speed-limiting computational resource which cannot be made significantly more efficient with an ASIC \cite{equihash}, \cite{balloon}, \cite{scrypt}. 
For example, many memory-bound hash functions have been implemented as PoW functions under the assumption that the price of memory units is roughly constant between custom and generalized hardware. 
Unfortunately, ASICs can still achieve modest performance gains because the energy required to power memory units in an ASIC is much lower than that of generalized hardware \cite{ren:bandwidth}. 
Additionally, because general purpose processors (GPPs) require significantly greater computational resources compared to the implementation of a specific function, any PoW function that utilizes only a subset of the resources within a GPP is vulnerable to an ASIC that mimics the GPP with respect to that subset and strips away everything else. 
Thus, we desire a PoW function for which a given GPP is {\em optimal} in die-area, resource allocation, and associated functionality, \textit{i.e.}, it stresses every computational resource in proportion with its importance within the GPP. 
If this is achievable, then any customized ASIC for such a PoW function would mimic the GPP, thus achieving our goal.

The goal of this work is to design a PoW function (which could also be called a ``hash'' function) that is most efficiently computed by a GPP such that no ASIC can be built for it that materially outperforms such a GPP. 
Stated another way, we assume that the GPP is a perfectly-optimized ASIC for an unknown set of programs. 
If a PoW function exists that requires execution of a wide ranging set of these unknown programs, then the GPP is an optimized-ASIC for that PoW function. 
From this viewpoint, we have {\em essentially reversed} the ASIC development thought process, which involves developing a custom chip that can execute a particular PoW function with high efficiency. 
In this work, we instead aim to develop a PoW function that can be executed on a GPP with optimal or near-optimal efficiency, resulting in a PoW system for which GPPs are the optimized ASIC.
We refer to this technique as {\em inverted benchmarking};  instead of designing hardware that can most efficiently execute a particular set of programs, we are designing a set of programs that can be most efficiently executed on a particular piece of hardware. 

{\bf HashCore} is a PoW function that satisfies the goal of inverted benchmarking, with an x86 processor serving as the working example. 
Note that we will henceforth use the term ``PoW function'' rather than ``hash function'' in this document, as this function is designed primarily to enable PoW and is not a conventional cryptographic hash function such as those in the SHA2 family. 
All other hashing and other functionality within the blockchain will remain the unchanged, with only the PoW function being replaced by HashCore.

At a high level, HashCore will consist of a sequence of code blocks called {\em widgets} that each stress one or more computational resources within the GPP. 
However, simply stressing each resource is not sufficient, as there is no guarantee that a program that arbitrarily stresses the computational structures in a chip cannot be executed more efficiently on another chip architecture. 
In particular, HashCore must near-optimally stress each resource so that no significant greater optimization is possible.
With HashCore, we take advantage of existing work in x86 CPU optimization by modeling our widgets after benchmarks that all x86 chips are optimized against, most notably the SPEC CPU 2017 benchmark suite. In this way, HashCore can leverage the research and development performed by all x86 chip designers. 

\color{black}
We reason that, as x86 chip manufacturers specifically design their hardware to execute such benchmarks efficiently, x86 processors are essentially near-optimal ASICs for such benchmarks\footnote{Note the emphasis on near-optimal. Even though there may be no optimality guarantee for such benchmarks, if no significant further optimization is possible, then using a custom ASIC over the GPP is rendered economically unviable.}.
As such, each of the computational structures present on x86 chips are there because they improve performance on SPEC CPU 2017 in some way. 
By designing widgets that emulate SPEC CPU 2017 workloads, we transitively ensure that x86 chips are near-optimal ASICs for HashCore. 

By a similar argument, if a chip is designed that consistently and significantly outperforms an x86 on HashCore, this chip should also significantly outperform an x86 on the SPEC CPU 2017 benchmark suite. 
Such a chip would be difficult to design since the costs associated with designing x86 CPUs are generally much higher than ASICs for a particular PoW function.
Not only would this newly designed ASIC be expensive, it would also be, by definition, a GPP capable of running a wide variety of x86 programs.
If such a chip were indeed designed, it would quickly replace existing x86 chips in the GPP market and be used for many more applications than just mining.

Note that, for ease of discussion and analysis, we speak of all x86 chips as if they share a single architecture. 
We reason that if there is some subset of x86 chips that are most performant on HashCore, then one or more of these chips will be readily available to the average user. 
We also believe that the performance gap will be small enough that there will exist a broad set of x86 chips that are competitive in the mining market, though we leave confirmation of this to future work.

\section{Related Work}
\label{sec:related}

``ASIC resistance'' has often been a desired goal for PoW functions in recent literature. 
Generally, a PoW function could be said to be ASIC resistant if significant performance gains could not be realized with custom hardware.
We have yet to see a formal definition of ASIC resistance, however, so we refer the reader to \cite{ren:bandwidth} and citations within for a more detailed description.

Hash functions such as Equihash \cite{equihash}, Balloon \cite{balloon}, and scrypt \cite{scrypt} all attempt to achieve ASIC resistance through memory-hardness. 
A memory-hard function is one that requires a large amount of memory to be executed efficiently, commonly requiring the traversal of a very large (potentially multiple gigabytes) graph. 
This property is desirable because upfront memory costs are roughly fixed between custom and general hardware, so higher memory requirements will reduce the price disparity between custom and general hardware in terms of hash rate per dollar spent. 
Unfortunately, this strategy does not take into account the cost of electricity to power the chip, which represents a large portion of operating expenses for miners. 
Ren et al. propose bandwidth-hard functions \cite{ren:bandwidth} in order to reduce the energy advantage of ASICs, though this still allows ASICs that consist only of many memory units and graph traversal logic units to achieve efficiency gains over a general purpose machine. 

We approach the problem similarly in this work, though rather than targeting a single computational resource to be stressed, we target every computational resource on a chip.
In other words, rather than attempting to design an ASIC resistant PoW function, we instead seek to design a PoW function for which a highly-optimized ASIC, such as an x86 processor, already exists and is widely available to consumers. 
This approach makes sense in the context of PoW systems for a variety of reasons, which will be discussed later, but it also allows us to take advantage of the considerable amount of research done towards optimizing general purpose CPUs.
Specifically, we assume that any general purpose CPU is intended to be an ASIC for the benchmark suites against which it is optimized. 
SPEC CPU 2017 is currently the primary benchmark suite against which x86 chips are optimized and is thus our focus in this work. 
SPEC CPU 2017 consists of 43 benchmarks that are drawn from end-user applications. 
Unfortunately, executing SPEC CPU 2017 in its entirety takes several hours for even the most performant CPUs, meaning HashCore can hope, at best, to simulate a small fraction of the actual workload during each hash computation. We overcome this problem by pseudo-randomly generating widgets that mimic the execution profile of SPEC CPU 2017. 

\section{Motivation}
\label{sec:motivation}

In this paper we encourage a perspective shift in the development of PoW functions with respect to ASIC resistance, as there is nothing inherently evil about ASICs.
On the contrary, optimized ASICs should be used by \emph{all} miners, which is only possible if they are broadly available to many users for roughly the same price. 
This is because the security of a PoW system depends on the property that computing a hash (a single execution of the PoW function) has roughly the same cost for all miners. 
Without this property, those miners who could compute hashes at lower cost would have a disproportionate advantage over the rest of the network and would therefore be able to write blocks at lower average cost than the remaining miners. 
This is clearly undesirable, as a ``perfectly" decentralized blockchain would not favor some set of miners over another. 
Then in an ideal PoW system, all miners would have equal access to the most efficient ASICs available at the same price. 
From this viewpoint, the desired invariant is not necessarily to resist the development of ASICs but rather to grant all miners equal opportunity to acquire and operate comparable, competitively efficient computational hardware. 
This opportunity is currently gated by the high costs and development times for near-optimal ASICs for a given PoW function. 
To this end, we develop HashCore, aimed to be a PoW function for which a GPP is already a near-optimal ASIC. 

There are many potential benefits to targeting GPPs as desired ASICs for a particular PoW function. 
Most notably, many individuals already own at least one GPP, meaning any would-be miner would not have to incur a significant upfront cost to enter the mining market. 
This would enable low-capital users to mine in a PoW system without disenfranchising larger miners, who can still purchase large quantities of GPPs and mine at a scale proportional to their means. 

We also observe that once a PoW function is determined to be effectively reducible to an ASIC, entities that are able to quickly develop and produce efficient ASICs may set prices significantly above the competitive equilibrium. 
These market inefficiencies manifest as additional costs on some miners (those who are not the developers), resulting in higher entry costs into the mining market until the market stabilizes.
However, as the GPP market is already highly competitive, there is far less potential for short-term pricing markups on GPPs. 
By targeting GPPs as the desired ASICs for HashCore, we can leverage a mature and efficient market for ASICs without these intermediate inefficiencies. 

Additionally, there is far less risk associated with investing in GPPs for cryptocurrency mining. 
Specialized ASICs, such as those used for Bitcoin, have very little use outside of mining. 
This means that during periods in which mining is less profitable, miners with specialized hardware may have to choose between mining at a potential loss or letting their machines sit idle. 
This would not be the case if general purpose hardware could be used to mine efficiently, as it by definition has value outside of its use in mining. 
Instead of sitting idle, mining hardware could be repurposed for alternate workloads or leased for cloud computing, for example. 
This reduction in risk may serve to encourage more users to begin mining, which will in turn strengthen the PoW system as a whole.

\section{Architecture}
\label{sec:arch}

\begin{figure*}[htbp]
\centerline{\includegraphics[width=6.5in,height=1.4in]{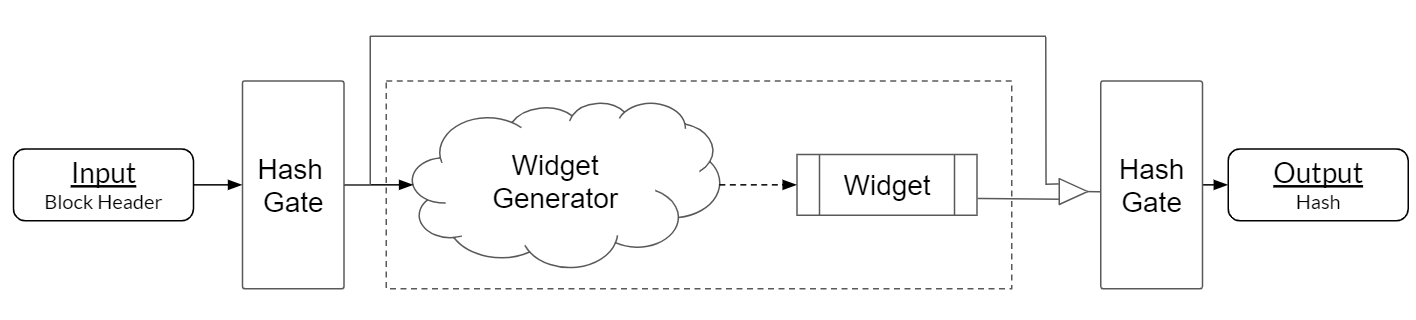}}
\caption{HashCore block diagram. The input is passed from left to right through a sequence of hash gates and widgets before the final output is realized. Each hash gate or widget will update the value it receives as input before passing it along to the next block. Propagation of data is denoted by solid arrows, concatenation of the Widget output and hash seed is denoted by the triangle, and the widget generation mechanism is denoted by a dashed arrow.}
\label{fig:HashCore}
\end{figure*}

Every execution of HashCore will consist of a sequence of code blocks that each affect the final output in some way. 
There are two types of code blocks in HashCore: a code block that executes a cryptographic hash function such as SHA2 will be referred to as a \emph{hash gate} and a code block that executes a workload designed to stress computational resources in an x86 chip will be referred to as a \emph{widget}. 
The basic HashCore algorithm can be observed in Figure \ref{fig:HashCore}. Initially, the input to HashCore will be passed through a hash gate, resulting in a fixed-size hash output that will be used twice, once as a random seed (which we will henceforth refer to as the \emph{hash seed}) for generating a widget and then again as part of the input to the final hash gate. 
The widget generated will, throughout execution, produce an output of variable size that is concatenated with the hash seed and passed through a second hash gate to achieve the final result.

Hash gates are used here to provide a simple mechanism through which HashCore can achieve the required cryptographic properties of pre-image resistance, second pre-image resistance, and collision resistance. 
We provide a formal proof of collision resistance in section V-A, which leverages the fact that the hash seed makes up part of the input to the second hash gate, allowing us to achieve collision-resistance regardless of the widget output. 
Additionally, so long as the hash gate is pre-image resistant, it will be nearly impossible ``select" a particular widget instantiation by constructing an input that targets a desired hash seed. 
Hash gates are also convenient for their ability to map a variable length input to a fixed length output; in our implementation we assume each hash gate produces a 256-bit output, though this is not required. 
Note that although we only make use of only a single widget in HashCore, it is certainly possible that multiple widgets could be generated for a given input string and executed sequentially.

\subsection{Widget Requirements}
Widgets represent the main computational task within each execution of HashCore.
At a high level, widgets must be short x86 programs that can be run most optimally on an x86 chip. 
Later we will describe how we achieve this property, but first we must introduce formal requirements for widgets that can be executed by HashCore:

\begin{itemize}
    \item \textbf{x86 Chip Utilization:} Each set of widgets must heavily utilize an x86 chip in such a way that, on average, no other chip architecture will materially outperform the most efficient x86 chip. 
    The main structures we are targeting most notably include the pipeline, branch predictor and re-order buffer, floating point units, arithmetic logic units, data and instruction caches, off-chip memory units, vector processing units, and the full x86 ISA. 
    As we will describe in the next section, we intend to target these structures through inverted benchmarking. 
    \item \textbf{Code Randomization:} Thus far, we have only discussed x86 chip utilization, but it is important to recognize that for any fixed program, custom hardware can be built that will materially outperform general purpose hardware. 
    This makes sense intuitively because, while general purpose machines must include functionality to perform well on a broad class of user programs, a custom machine can be precisely engineered to a particular program. 
    SPEC CPU 2017 workloads are somewhat robust against this problem because they are long enough that attempting to hand-engineer custom hardware for them is infeasible. 
    Unfortunately, each execution of HashCore should have running time on the order of a few seconds or less in order to accommodate sub-minute block times like those of Ethereum \cite{ethereum}, meaning we must solve this problem in another way. 
    By randomizing the widgets to be executed for each distinct input to HashCore, we ensure that no ASIC can optimize for a particular code path. 
    
    Randomization could be achieved in two ways within our framework. 
    We could either fix a large set of widgets for HashCore and select an ordered subset of them based on the output of the first hash gate, or we could use the output of the first hash gate as a random seed to generate a widget or set of widgets at run-time. 
    We opt for the latter option, which provides the additional property that the machine running HashCore must be able to run the widget generation script. 
    This does not limit general purpose machines, which can easily accomplish such a task, but may make designing specialized hardware more challenging. 
    \item \textbf{Irreducible Widgets}: 
The widget should also be irreducible in the sense that certain code segments cannot be skipped and the output cannot be predicted without full execution of the widget.
If a widget could be reduced to an easier piece of code with minimal effort, then the benefits of HashCore would be lost.
In that event, any individual access to the reduction algorithm would be able to gain an unfair advantage over the rest of the mining pool.
Worse, the x86 optimal nature of HashCore would likely be lost if a widget could be reduced to simpler code. 
For these reasons, it is critical that a widget be is not reducible to a simpler program in polynomial time.
\end{itemize}

\subsection{Widget Generation}
Widgets are generated at run-time through the inverted benchmarking technique. 
Inverted benchmarking consists of generating widgets to match the execution profile of an existing CPU benchmark, such as SPEC CPU 2017\cite{SPEC}, such that the resulting widgets run optimally on a GPP.

The inverted benchmarking technique builds on previous research in the area of proxy generation \cite{perfprox}. 
The goal of proxies is to generate short running programs that mimic the behavior of much longer programs. 
Computer architects use these proxies to test new computer designs on simulators that run at much slower speeds than silicon hardware. 
A key requirement is that the proxy must stress a design almost identically to how the original program would have.

Proxies overlap well with HashCore's requirements after a few minor adjustments.
Proxies heavily require matching performance of the original workload, while HashCore requires a large variety of pseudo-random workloads. 
For this reason, HashCore modifies an existing proxy technique, PerfProx \cite{perfprox}, to achieve these goals.

Before discussing the modifications further, we need to first understand the basic details of Perfprox \cite{perfprox}. 
Perfprox functions by profiling a selected workload on a variety of performance metrics such as instruction mix, branch behavior, memory access patterns, and data dependencies. Using this performance profile, Perfprox then creates a synthetic workload that has nearly identical performance metrics. 
Please refer to \cite{perfprox} for a more detailed discussion. 
This framework provides an adequate foundation of producing varied x86 programs.

HashCore modifies the Perfprox technique in two key ways. 
First, the random seed must be added to the performance profile as an input. 
The seed is distributed among various performance metrics as outlined in Table \ref{table:seed}. 
The 256-bit seed is divided into eight 32-bit integers that are added to the  performance profile. 
The exception to this are the last two 32-bit values which are used to seed pseudo-random number generators. 
This means that each seed will add some amount of noise to the widget generator so that each widget has slightly different performance, resulting in a distribution of widgets centered around the target performance profile. 
While this added noise would be unacceptable for a proxy, it is necessary for HashCore in order to fulfill the code randomization requirement discussed above.

\begin{table}[htbp]
\caption{Hash Seed Usage}
\begin{center}
\begin{tabular}{|c|c|}
\hline
Hash Bits& Usage \\
\hline
0-31 & Integer ALU \\
32-63 & Integer Multiply \\
64-95 & Floating Point ALU \\
96-127 & Loads \\
128-159 & Stores \\
160-191 & Branch Behavior \\
192-223 & Basic Block Vector Seed \\
224-255 & Memory Seed \\
\hline
\end{tabular}
\label{table:seed}
\end{center}
\end{table}

Second, HashCore modifies the proxies generated so that they produce an output. 
There is no need for traditional proxies to produce an output, but HashCore requires that each widget produce a unique signature dependent on the complete execution of the widget. 
This problem is easily solved by forcing the proxy to output register values throughout execution, resulting in an output string composed of the sequence of register states at various points in execution. 
This forces the entire workload to be executed because every x86 instruction modifies the registers in some way, and if even a single bit is incorrect in the proxy output then the resulting hash will be invalid.

Another property of utilizing the Perfprox technique is the additional execution requirements it places on HashCore.
Three programs must be run to produce the widget's output. Perfprox first runs a python script that takes the performance profile and hash seed as an input and produces a C program. 
Next, the GCC compiler compiles the C program into an executable x86 program. 
Finally, the x86 program itself must be run and the output recorded. 
This widget generation process is not limiting for a typical x86 system, though it may increase the difficulty of developing custom hardware for HashCore, depending on the approach. 

\section{Results}
\label{sec:results}

This section covers a proof of collision-resistance and some initial experiments done on HashCore to demonstrate its utility. 
We first show that, assuming the hash gates used in our architecture are collision-resistance, we can achieve collision-resistance on HashCore regardless of the functionality of the widget generation and execution procedures.
This is a required result for any PoW function, as any collisions found could threaten the tamper-resistance of the underlying blockchain being secured.  
Next we analyze the performance metrics of the widgets to show that similar performance values are produced in comparison to the original workload. 
Ideally, a set of widgets should produce a spread of performance values centered around the original workload's value.  

All experiments were done on a set of one thousand widgets generated on a state-of-the-art platform, a Dell PowerEdge R320 server equipped Xeon E5-2430 v2 processor (codenamed Ivy Bridge), and 64 GB DDR3 memory. 
Each single-threaded widget was run on a single core of this system with minimal overhead from the operating system due to core isolation. 
For these experiments, we profiled a single workload from the SPEC CPU 2017 benchmark suite, namely the Leela integer speed workload. 
However, there is nothing unique about this workload, and similar widgets could be produced for a variety of workload performance profiles. 

The widgets under test were produced by randomly generating one thousand hash seeds and combining those seeds with the Leela performance profile in the manner described above. 
These widgets produced outputs ranging in size from 20 kilobytes to 38 kilobytes with a large amount of variation in register contents during execution. In more detail, the output is a series of snapshots of the computer's register contents captured every few thousand instructions. These instructions sequentially modify the register state, which leads to a dependency both on the initial register state and the instructions executed.  

\subsection{Proof of Collision-Resistance}
In this section we describe our proof of the collision-resistance of HashCore under the assumption that the hash gates being used are collision-resistant. More concretely, we will prove Theorem 1 using a fairly straightforward reduction, a proof technique that involves reducing the security guarantee of a system to that of a primitive within the system. 
First, we will describe some of the definitions used in the proof at a high level, though for more complete definitions we refer the reader to \cite{katz-lindell}. 

A \emph{probabilistic polynomial time (PPT)} attacker is one who has the ability generate truly random bits and is allowed to make a polynomial (in some security parameter specific to the security scheme in question) number of queries. 
In this paper, one query is a single execution of HashCore. 
Requiring the adversary to be computationally bounded is crucial to our analysis, as collisions exist by definition in any compressing hash function, and an unbounded adversary could simply enumerate every input-output pair and find a collision with at least the same advantage. 
A function $f(x)$ is said to be \emph{negligible} if it grows slower than any $\frac{1}{x^c}$ for any constant $c$. 
Finally, a \emph{Collision-Resistant Hash Function (CRHF)} is a hash function with the following security property: given a full description of the function, any PPT adversary can find a collision with at most negligible probability. 
Put another way, it should be so improbable that any PPT adversary can find a collision on a CRHF that it will essentially never happen.
In this work, we make the fairly standard assumption that SHA-256, the hash function we use in our hash gates, is a CRHF. 

Let $G: \{0, 1\}^* \to \{0, 1\}^n$ be a CRHF, $W: \{0, 1\}^n \to \{0, 1\}^l$ be a function that maps $n$ bits to $l$ bits, and $H: \{0, 1\}^* \to \{0, 1\}^n$ be a function that uses $G$ and $W$ as intermediate steps to compute its end result. 
Specifically, for an input $x$, let $s=G(x)$ be the hash seed generated from the first hash gate. Then $H(x)=G(s||W(s))$ fully defines the function of HashCore, where $||$ denotes the concatenation of two bit-strings, $G$ denotes a single hash gate, and $W$ denotes the complete widget generation and execution process. 

\newtheorem{crhf}{Theorem}

\begin{crhf}[Collision-Resistance of HashCore]
$H$ is a CRHF if $G$ is a CRHF.
\end{crhf}

We will provide an informal overview of the proof of Theorem 1 here; for a more formal proof we direct the reader to the appendix. 
Intuitively, the proof proceeds in the following way. 
We first assume that $H$ is not a CRHF, meaning there exists a PPT adversary that is able to find a collision on HashCore. 
We then show that if such an adversary exists, we can use this adversary to find a collision on $G$ with probability 1, as any collision on $H$ will require a collision on either the first or second execution of $G$. 
This leads to a contradiction, however, as we have assumed $G$ to be a CRHF. 
It follows that if $G$ is a CRHF, then $H$ must also be a CRHF.

We also remark that although we denote the output of $W$ to be an $l$-bit string in the proof for ease of understanding, $W$ actually has variable-length output in our implementation of HashCore.
This does not affect the validity of the proof, as all of the analysis performed is agnostic to the output size of $W$; in fact, the result holds regardless of any property of $W$ so long as $W$ can be executed in polynomial time\footnote{In order to avoid instantiations of $W$ that simply brute force search for a collision on $G$, for example.}.
This allows us to achieve collision-resistance essentially for free, strengthening the HashCore framework for future work.

\subsection{Performance Experiments}
This section outlines our analysis of the widgets with respect to how they matched the Leela workload performance values. While not a strict requirement for HashCore, the widgets should have similar performance characteristics to Leela in order to apply the inverted benchmarking idea to a CPU. 
With this in mind, we measured various performance metrics for the thousand sample widgets and compared them to the original workload.

\begin{figure}[t]
  \centering
  \label{fig:ipc}
  \includegraphics[height=0.2\textheight,width=1\linewidth]{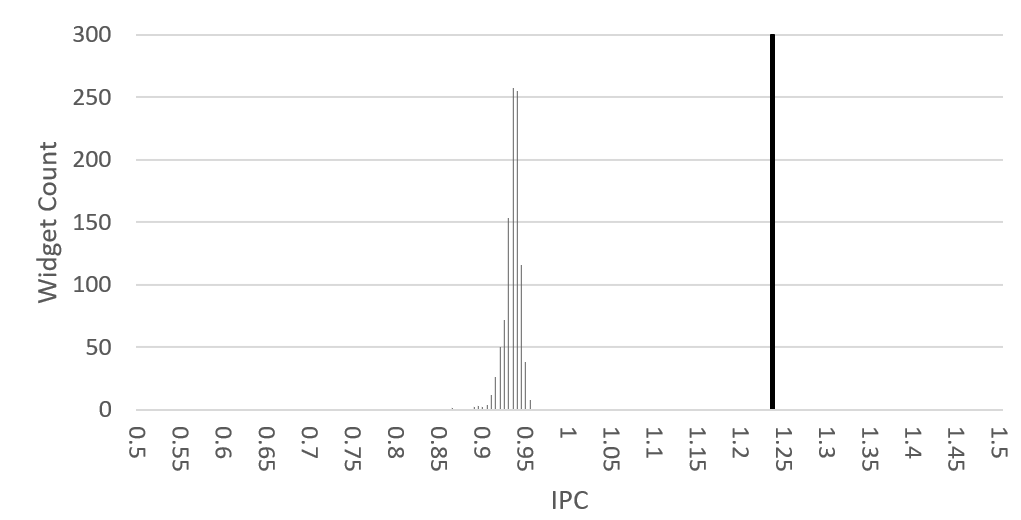} 
  \caption{IPC Widget Comparison}
  \vspace{-4mm}
\end{figure}

\begin{figure}[t]
  \centering
  \label{fig:brpr}
  \includegraphics[height=0.2\textheight,width=1\linewidth]{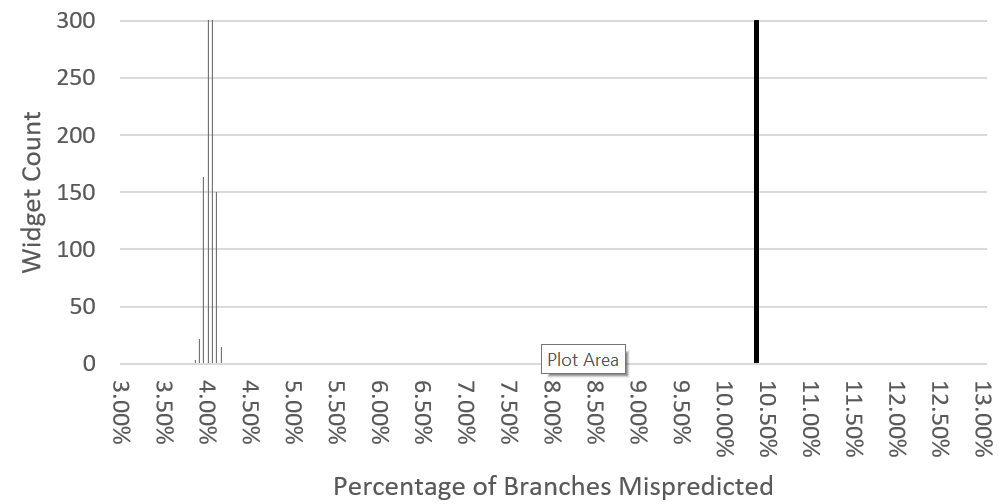} 
  \caption{Branch Prediction Widget Comparison}
  \vspace{-4mm}
\end{figure}

The performance results are outlined in Figure \ref{fig:ipc}. 
As can be seen for IPC, the widgets seem to follow a roughly Gaussian distribution with a mean slightly lower than those of the original Leela workload. 
Such a shift in performance is acceptable for our purposes and helps highlight the noise being added by the hash seed.
Importantly, HashCore only adds positive noise to the instruction type counts.
This increase in instructions leads to proportionally less branch instructions which are often a limiting factor to performance. The branch behavior further solidifies these results as shown in Figure \ref{fig:brpr}. These results are expected given how the widget generator was designed.

\section{Discussion}
\label{sec:discussion}

\subsection{Widget Generation vs Selection}
In this work we generate widgets randomly at runtime to achieve sufficiently random code, though we believe that storing a fixed set of widgets instead could be a viable alternative.
This kind of solution would likely require a very large widget pool, perhaps consisting of as many total lines of code as SPEC CPU 2017 or more, in order to discourage chip designers from engineering ASICs specifically for each widget. 
Then each execution of the PoW function would consist of gating the input string and using the result to select some ordered set of these widgets to be executed, resulting in an output string to be hashed. 

While random generation and random selection seem similar on the surface, there are several important tradeoffs to consider.
In terms of storage, random generation likely wins out, as the widget pool used for selection could consist of several gigabytes worth of code depending on the implementation.
Selection also comes with the risk that custom ASICs could be constructed for some subset of the widget pool, resulting in modest performance gains any time these widgets are selected.
On the other hand, widget selection is far less computationally intensive than widget generation no matter how efficient the generator script is. 
This would mean that widget execution would account for a higher percentage of the total execution time of the PoW function, which in turn could mean greater utilization of the GPP in question. 
Additionally, targeting and analyzing the resource utilization of the class of widgets produced by a generator script may be more difficult than it would be for a pre-determined pool of widgets. 
We believe it is possible that some combination of both approaches could yield a robust PoW function, though we leave this to future work. 

\subsection{Targeting alternative GPPs}
Although we target x86 ICs in this work, there is no reason that the HashCore framework could not be leveraged on a variety of other chip architectures, such as ARM cores or GPUs. 
We target x86 chips because most would-be miners already own x86 chips in their personal computers and most personal computers can easily run mining software, though it would be interesting to consider constructing a PoW function optimized for the ARM cores in mobile phones, for example. 
We have intentionally made HashCore fully modular, so modifying HashCore to target alternate architectures would require only that a new widget pool or widget generator script be developed. 

\subsection{Alternatives to Inverted Benchmarking}
Our motivation for using inverted benchmarking, which for HashCore simply involves constructing widgets that mimic the execution pattern of SPEC CPU 2017, is based on the assumption that all x86 ICs are ASICs for SPEC CPU 2017. 
We seek to leverage the research and development performed by x86 architects in our widget design. 
However, alternate strategies exist, such as that of RandomX \cite{randomx}.
The developers of RandomX approach the problem from a different angle by constructing a virtual machine that attempts to simulate a generic GPP. 
Their strategy involves generating a random program to fit into the VM they define before executing it, followed by a hash on the output. 
This differs from HashCore only in the program generation methodology, where we target a particular execution profile matching SPEC CPU 2017 and they instead target explicit utilization of each computational structure. 
It could be that some combination of these two approaches would be effective, though we also leave this for future work to determine.

\subsection{Repurposing Mining Hardware}
In addition to reducing the pricing and availability barriers against entering the mining market, targeting GPPs as ASICs for a PoW function comes with an additional benefit when we consider that the revenue available in mining markets varies over time. 
One problem we see with mining today is that ASICs for current PoW functions have little use outside of mining. 
By constructing a PoW function that can be efficiently mined with a GPP, we allow mining hardware to be repurposed as needed. 
It may be profitable in some mining markets for users to mine with their personal computer over night, for example. 
More importantly, as fluctuations in transaction volume or market capitalization affect the profitability of mining a cryptocurrency over time, large-scale miners often must decrease the amount of hashing power they contribute to a PoW system in order to maximize profits. 
In the absence of alternative mining markets, these machines could easily be repurposed for a time as cloud compute servers, for example. 
Because GPPs have value outside of mining, we can reduce some of the risks associated with purchasing and owning mining hardware, further reducing the barriers to entering the mining market. 

\subsection{Additional Benefits of a Modular PoW Function}
Another interesting consideration is that the HashCore framework is agnostic to the code executed by each widget so long as it meets the requirements detailed in Section IV. 
One of the most common criticisms of PoW systems today is that, because the search for suitable nonces has no external benefits, mining appears to be a waste of energy. 
This is not so, as energy is being expended to guarantee immutability of the blockchain to some degree. 
Still, it is worth considering that philanthropic or otherwise useful workloads could be injected as widgets into the HashCore framework to improve public perception and provide non-economic incentives to potential miners. 
For example, there may be a way that tasks that require high computational effort, such as protein folding \cite{folding} or searching for extraterrestrial life \cite{seti}, could be baked into the widgets used in a PoW function. 
It is worth noting that this has the potential to degrade the security guarantees of a PoW system by providing malicious actors with additional incentives to attack. 
We do not believe this will be the case, as the additional incentives are shared by both honest and malicious miners, but we leave this security analysis again to future work. 

\subsection{Future Work}
While we believe that HashCore is a step in the right direction towards a completely decentralized mining market, there are still a few open questions that must be worked out before HashCore can be viable in a real PoW system. 
Most notably, obtaining provable guarantees that a set of workloads are most efficiently run on a particular chip design is an open question for computer architects. 
Instead, we currently rely on a heuristic argument that an x86 IC is a near-optimal ASIC for SPEC CPU 2017 and that HashCore mimics SPEC CPU 2017 in execution pattern. 
At a higher level, we also observe that HashCore alone will not guarantee a decentralized mining market in the long term. 
At a competitive equilibrium, economies of scale and access to cheap energy will likely result in a relatively small number of large mining operations located wherever the cheapest electricity can be purchased. 
Even so, HashCore will still reduce entry barriers before a competitive equilibrium is reached and allow for hardware repurposing and the potential for philanthropic workloads at market equilibrium.

\section{Conclusions}
\label{sec:conclusion}
The primary motivation for developing HashCore is to promote a more accessible and competitive mining market for cryptocurrencies utilizing a Proof-of-Work protocol to achieve blockchain consensus. 
To this end, we have identified an ASIC that many would-be miners already own, an x86 CPU, and have developed a PoW function to match this ASIC. 
HashCore consists of a pseudo-randomly generated x86 program called a widget that must be executed in its entirety in order to achieve a valid result. 
By generating widgets that mimic the execution profile of SPEC CPU 2017, a workload for which x86 chips can be thought of as ASICs, we are able to take advantage of the research and development performed by x86 chip developers.
This work represents an inversion of the problems faced by both x86 chip developers and ASIC developers, as both of these groups seek to optimize hardware against a fixed workload while we seek to optimize a workload against fixed hardware. 
We believe that this approach will reduce the barriers for users to profitably mine cryptocurrencies and facilitate less centralized mining markets.

\section*{Acknowledgments}
We would like thank all researchers at The University of Texas at Austin who assisted with this research. In particular, we thank Dr. Karl Kreder and Dr. Mohit Tiwari for the insightful discussions we had about the overarching goals and requirements of HashCore. We also thank Dr. Brent Waters for his insights related to the collision-resistance of HashCore.

\newpage
\appendix[Proof of Collision-Resistance of HashCore]
\label{sec:proof}
For readability, we repeat the definitions from Section V here. 

Let $G: \{0, 1\}^* \to \{0, 1\}^n$ be a CRHF, $W: \{0, 1\}^n \to \{0, 1\}^l$ be a function that maps $n$ bits to $l$ bits, and $H: \{0, 1\}^* \to \{0, 1\}^n$ be a function that uses $G$ and $W$ as intermediate steps to compute its end result. 
Specifically, for an input $x$, let $s=G(x)$ be the hash seed generated from the first hash gate. 
Then $H(x)=G(s||W(s))$, where $||$ denotes the concatenation of two bit-strings, fully defines HashCore. 

\newtheorem{crhf2}{Theorem}

\begin{crhf2}[Collision-Resistance of HashCore]
$H$ is a CRHF if $G$ is a CRHF.
\end{crhf2}

\begin{proof}
Assume that there exists a PPT adversary $\mathcal{A}$ that can find $x_0, x_1 \in \{0,1\}^*$ such that $x_0 \neq x_1$ and $H(x_0) = H(x_1)$ with non-negligible advantage, defined as $\mathsf{adv}_{\mathcal{A}} = $ Pr[$\mathcal{A}$ finds a collision on $H$]. We will construct a PPT algorithm $\mathcal{B}$ that uses $\mathcal{A}$ to break the CRHF security of $G$. $\mathcal{B}$ is defined below. 

\begin{enumerate}
    \item $\mathcal{B}$ receives the function $G$ from the CRHF challenger and constructs $H$. $\mathcal{B}$ sends a full description of $H$ to $\mathcal{A}$. 
    \item Next, $\mathcal{B}$ receives $\hat{x}_0, \hat{x}_1$ from $\mathcal{A}$ as an attempt at a collision on $H$. $\mathcal{B}$ constructs $x_0^*, x_1^*$ in the following way.
    \begin{itemize}
        \item If $H(\hat{x}_0) = H(\hat{x}_1)$ and $\hat{x}_0 \neq \hat{x}_1$, compute $s_0=G(\hat{x}_0)$, $s_1=G(\hat{x}_1)$. 
        \begin{itemize}
            \item If $s_0 = s_1$, set $x_0^* = \hat{x}_0$, $x_1^* = \hat{x}_1$. 
            \item Else, set $x_0^* = s_0||W(s_0)$,  $x_1^* = s_1||W(s_1)$.
        \end{itemize}
        
        \item Else, $\mathcal{B}$ selects $x_0^*, x_1^*$ at random.
    \end{itemize}
    
    Clearly, $\mathcal{A}$ wins if $H(\hat{x}_0) = H(\hat{x}_1)$ and $\mathcal{B}$ wins if $G(x_0^*) = G(x_1^*)$.
    
\end{enumerate}

Next, we will argue that if $\mathcal{A}$ finds a collision on $H$, then $\mathcal{B}$ is guaranteed to produce a collision on $G$ with the above algorithm. In other words, given $\hat{x}_0 \neq \hat{x}_1$ such that $H(\hat{x}_0) = H(\hat{x}_1)$, $\mathcal{B}$ can find $x_0^* \neq x_1^*$ such that $G(x_0^*) = G(x_1^*)$. We will argue this case by case.

\medskip
\noindent\textbf{Case 1}: $s_0 = s_1$.

Clearly in this case, $x_0^* = \hat{x}_0, x_1^* = \hat{x}_1$ is a collision on $G$, as $s_0=G(\hat{x}_0)$, $s_1=G(\hat{x}_1)$.

\medskip
\noindent\textbf{Case 2}: $s_0 \neq s_1$.

In this case, we know that $s_0||W(s_0) \neq s_1||W(s_1)$ and that $G(s_0||W(s_0)) = G(s_1||W(s_1))$. Then $x_0^* = s_0||W(s_0), x_1^* = s_1||W(s_1)$ is a collision on $G$. 

\medskip
It is then clear that $\mathsf{adv}_{\mathcal{B}} = $ Pr[$\mathcal{B}$ finds a collision on $G$] $\geq$ Pr[$\mathcal{A}$ finds a collision on $H$] = $\mathsf{adv}_{\mathcal{A}}$. This leads to a contradiction, as we have assumed that $G$ is a CRHF. The statement follows.

\end{proof}



\begin{thebibliography}{00}

\bibitem{equihash} A. Biryukov and D. Khovratovich, ``Equihash: Asymmetric Proof-of-Work Based on the Generalized Birthday Problem,'' Ledger, [S.l.], vol. 2, pp.1--30

\bibitem{balloon} D. Boneh, H. Corrigan-Gibbs, and S. Schechter, ``Balloon Hashing: A Memory-Hard Function Providing Provable Protection Against Sequential Attacks,'' in Advances in Cryptology -- ASIACRYPT 2016. Lecture Notes in Computer Science, vol 10031. Springer, Berlin, Heidelberg

\bibitem{pow:dwork} C. Dwork and M. Naor, ``Pricing via Processing or Combatting Junk Mail,'' in Advances in Cryptology --- CRYPTO' 92. Lecture Notes in Computer Science, vol 740. Springer, Berlin, Heidelberg

\bibitem{katz-lindell} J. Katz and Y. Lindell, Introduction to Modern Cryptography, Chapman and Hall/CRC, Boca Raton,
FL, 2008.

\bibitem{folding} ``Folding@home.'' https://foldingathome.org/

\bibitem{pow} M. Jakobsson, A. Juels, ``Proofs of Work and Bread Pudding Protocols(Extended Abstract),'' in Secure Information Networks. IFIP — The International Federation for Information Processing, vol 23. Springer, Boston, MA

\bibitem{bitcoin} S. Nakamoto, ``Bitcoin: A Peer-to-Peer Electronic Cash System,'' unpublished.

\bibitem{perfprox}R. Panda and Lizy Kurian John, “Proxy benchmarks for emerging big-data workloads”,  in 2017 IEEE International Symposium on Performance Analysis of Systems and Software (ISPASS’17), Santa Rosa, CA, USA.

\bibitem{scrypt} C. Percival,  ``Stronger key derivation via sequential memory-hard functions,'' self-published.

\bibitem{ren:bandwidth} L. Ren and S. Devadas, ``Bandwidth Hard Functions for ASIC Resistance,'' Cryptology ePrint Archive, Report 2017/225

\bibitem{seti} ``Seti@home.'' https://setiathome.berkeley.edu/

\bibitem{SPEC}“SPEC CPU2017.” https://www.spec.org/cpu2017.

\bibitem{randomx} Tevador, ``RandomX: Experimental proof of work algorithm based on random code execution.'' Github repository. https://github.com/tevador/RandomX

\bibitem{ethereum} G. Wood, ``ETHEREUM: A SECURE DECENTRALISED GENERALISED TRANSACTION LEDGER,'' self-published. 

\end{thebibliography}
\end{document}